\documentclass{emulateapj}
\usepackage{apjfonts}
\lefthead{HAN}
\righthead{Microlensing Exo-satellite Signals}

\def\eqalign#1{\null\,\vcenter{\openup\jot
        \ialign{\strut\hfil$\displaystyle{##}$&$
        \displaystyle{{}##}$\hfil \crcr#1\crcr}}\,}

\lefthead{HAN} 
\righthead{MICROLENSING MOONS}

\begin{document}
\title{Microlensing Detections of Moons of Exoplanets}

\author{Cheongho Han}
\affil{Program of Brain Korea 21, Department of Physics, 
Chungbuk National University, Chongju 361-763, Korea;\\
cheongho@astroph.chungbuk.ac.kr}

\submitted{Submitted to The Astrophysical Journal}

\begin{abstract}
We investigate the characteristic of microlensing signals of 
Earth-like moons orbiting ice-giant planets.  From this, we find 
that non-negligible satellite signals occur when the planet-moon 
separation is similar to or greater than the Einstein radius of 
the planet.  We find that the satellite signal does not diminish 
with the increase of the planet-moon separation beyond the Einstein 
radius of the planet unlike the planetary signal which vanishes 
when the planet is located well beyond the Einstein radius of the 
star.  We also find that the satellite signal tends to have the 
same sign as that of the planetary signal.  These tendencies are 
caused by the lensing effect of the star on the moon in addition 
to the effect of the planet.  We determine the range of satellite 
separations where the microlensing technique is optimized for the 
detections of moons.  By setting an upper limit as the angle-average
of the projected Hill radius and a lower limit as the half of the 
Einstein radius of the planet, we find that the microlensing method 
would be sensitive to moons with  projected separations from the 
planet of $0.05\ {\rm AU} \lesssim d_{\rm p} \lesssim 0.24\ {\rm AU}$ 
for a Jupiter-mass planet, $0.03\ {\rm AU}\lesssim d_{\rm p} \lesssim 
0.17\ {\rm AU}$ for a Saturn-mass planet, and $0.01\ {\rm AU} \lesssim 
d_{\rm p} \lesssim 0.08\ {\rm AU}$ for a Uranus-mass planet.  We 
compare the characteristics of the moons to be detected by the 
microlensing and transit techniques.
\end{abstract}



\section{Introduction}

All planets in our solar system except Mercury and Venus have 
moons.  With the increasing number of discovered extrasolar 
planets, the existence of moons and their characteristics in 
these exoplanets emerge as new questions.  Several methods to 
answer these questions have been proposed.  \citet{sartoretti99} 
pointed out that high-precision photometry of stars during planet 
transit can be used  to detect extrasolar moons either by direct 
satellite transit or perturbation in the timing of the planet 
transit.  \citet{brown01} applied this method to the transit 
planet HD 209458b and placed upper limits on moons orbiting the 
planet by using the transit light curve obtained from {\it Hubble 
Space Telescope} observations.

In addition to the transit method, microlensing technique can also 
be used for the detections of extrasolar moons.  This possibility 
was first mentioned by \citet{bennett02}.  They claimed that 
space-based lensing surveys with high precision and cadence would 
be able to detect not only planets but also moons orbiting the 
planets.  From the investigation of satellite-induced lensing 
signals, \citet{han02} pointed out that detections of Earth-Moon 
like systems would be difficult because the satellite signal would 
be seriously smeared out by severe finite source effect.  However, 
moons with large masses may exist.  From detailed investigation 
of the long-term dynamical stability of moons, \citet{barnes02} 
pointed out that Earth-like moons of Jovian planets could have 
stable orbits for  long time scales.  If such massive moons are 
common, it will be possible to detect them by using the microlensing 
technique.

In this paper, we investigate the characteristics of the lensing 
signals of Earth-like moons orbiting ice-giant planets.  We 
investigate the variation of satellite signals depending on the 
locations of satellites and masses and locations of host planets.  
We also determine the range of satellite separations where the 
microlensing technique is optimized for the detections of moons.

\section{Basics of Lensing}

For the description of the lensing behavior produced by satellite
systems, it is required to include at least three lens components 
of the host star, planet, and moon.  For a multiple-lens system, 
the image mapping from the lens plane to the source plane is 
expressed as 
\begin{equation}
\zeta = z - \sum_{k=1}^N {m_k/M \over \bar{z}-\bar{z}_{L,k}},
\label{eq1}
\end{equation}
where $N$ is the number of the lens components, $\zeta=\xi + i\eta$, 
$z_{L,k}=x_{L,k}+iy_{L,k}$, and $z=x+iy$ are the complex notations 
of the source, lens, and image positions, respectively, $\bar{z}$ 
denotes the complex conjugate of $z$, $m_k$ are the masses of the 
individual lens components, $M=\sum_k m_k$ is the total mass of 
the system, and $m_k/M$ represent the mass fractions of the individual 
lens components.  Here all lengths are expressed in units of the 
Einstein radius that is related to the lens mass and the distances 
to the lens ($D_{\rm L}$) and source ($D_{\rm S}$) by
\begin{equation}
\eqalign{
\theta_{\rm E}=
     & \left({4GM\over c^2}\right)^{1/2} 
       \left({1\over D_{\rm L}}-{1\over D_{\rm S}} \right)^{1/2} \cr
\sim & 0.55\ {\rm mas}\ 
       \left( {M\over 0.3\ M_\odot}\right)^{1/2}
       \left( {D_{\rm S}\over 8\ {\rm kpc}}\right)^{-1/2}
       \left( {D_{\rm S}\over D_{\rm L}}-1\right)^{1/2}.  \cr
}
\label{eq2}
\end{equation}
Due to lensing, the image of the source star is split into multiple 
fragments and the individual images are distorted.  The fragmentation 
and distortion of the source image cause variation of the source 
brightness.  The lensing process conserves the source surface 
brightness, and thus the magnification of each image corresponds 
to the ratio between the areas of the image and source. For an 
infinitesimally small source, the magnification of each image is 
obtained by the Jacobian of the mapping equation, i.e.\ 
\begin{equation}
A_i = \left\vert \left( 1-{\partial\zeta\over\partial\bar{z}}
{\overline{\partial\zeta}\over\partial\bar{z}} \right)^{-1} 
\right\vert.
\label{eq3}
\end{equation}
For Galactic lensing events, the typical separations between 
images are of the order of 0.1 mas and thus the individual images 
cannot be resolved.  However, events can be noticed by the variation 
of the source star flux where the total magnification corresponds 
to the sum of the magnifications of the individual images, i.e.\ 
$A=\sum_i A_i$.

One important difficulty in describing the lensing behavior of 
a multiple lens system is that the mapping equation is expressed 
in terms of the source position as a function of the image and 
lens positions.  This implies that finding image positions for a 
given source position requires inversion of the mapping equation 
but the inversion is algebraically impossible for a multiple lens 
system.  One way to obtain the image positions is expressing the 
mapping equation as a polynomial in $z$ and then numerically 
solving the polynomial \citep{witt95}.  The advantage of this 
method is that it enables semi-analytic description of the lensing 
behavior and saves computation time.  However, the order of 
polynomial increases as $N^2+1$ \citep{rhie97} and thus solving 
the polynomial becomes difficult as the number of lens components 
increases.  In this case, one can still obtain the magnification 
patterns by using the inverse ray-shooting technique \citep{schneider86, 
kayser86, wambsganss90}.  In this method, a large number of light 
rays are uniformly shot from the observer plane through the lens 
plane and then collected (binned) in the source plane.  Then, the 
magnification pattern is obtained by the ratio of the surface 
brightness (i.e., the number of rays per unit area) on the source 
plane to that on the observer plane.  Once the magnification pattern 
is constructed, the light curve resulting from a particular source 
trajectory corresponds to the one-dimensional cut through the 
constructed magnification pattern.  Although this method requires 
a large amount of computation time for the construction of detailed 
magnification patterns, it has an important advantage that the 
lensing behavior can be investigated regardless of the number 
of lenses.  In addition, one can easily incorporate the finite 
source effect, which is important for the description of the 
perturbations caused by low-mass objects such as planets and 
moons \citep{bennett96}.  Due to this reason, we use the 
ray-shooting method for the investigation of magnification 
patterns.

Due to the small mass ratio of the planet and even smaller mass 
of the moon, the lensing light curve of an event produced by a 
star having a planet with moons is well described by the 
single-lens light curve produced by the host star for most of 
the event duration.  A short-duration perturbation occurs when 
the planet happens to be at the location of one of the two images 
of the source star produced by the host star \citep{gaudi97}.  
Since the moon is close to the planet, the moon can also perturb 
the image and produce an additional anomaly.  The position of 
the image-perturbing planet in the lens plane corresponds to 
the position of caustic in the source plane.  In other words, 
perturbations occur when the source is located close to the 
caustic.  The caustic represents the set of positions in the 
source plane at which the magnification of a point source event 
is infinite.  For the binary lens case composed of the star and 
planet, there exist two sets of caustics.  One is located very 
close to the star (central caustic) and the other is located 
away from the star (planetary caustic).  Among the two perturbation 
regions around the individual caustics, noticeable perturbations 
induced by the moon are expected only in the region around the 
planetary caustic.  Two factors cause difficulties in finding 
satellite signatures around the central caustic region.  First, 
the central perturbation produced by the moon  occurs in a very 
tiny region.  As a result, the signature of the moon would be 
significantly washed out by the finite source effect.  Second, 
the perturbation regions of the planet and the moon nearly 
coincide.  Then, the anomaly in the lensing light curve would 
be dominated by that of the planet due to the overwhelming mass 
of the planet compared to the mass of the moon, making it even 
more difficult to identify the satellite signature.  We therefore 
focus on the perturbation region around the planetary caustic 
throughout the paper.

The location of the planetary caustic is related to the star-planet
separation by
\begin{equation}
{\bf s}_{\rm c}={\bf s}_{\rm p}\left( 1- {1\over s_{\rm p}} \right)^2,
\label{eq4}
\end{equation}
where ${\bf s}_{\rm p}$ represents the position vector of the planet 
from the star and its length is normalized by the Einstein radius of 
the star.  Then, the caustic is located on the planet side when the 
planet is outside the Einstein ring ($s_{\rm p}>1.0$), while it is 
located on the opposite side when the planet is inside the ring 
($s_{\rm p}<1.0$).  The number of caustics also depends on the 
planetary separation and it is one when $s_{\rm p}>1.0$ and two when 
$s_{\rm p}<1.0$.  The caustic is within the Einstein ring when the 
planetary separation is within the range of $0.6\lesssim s_{\rm p}
\lesssim 1.6$.  The caustic size, which is proportional to the 
chance of planetary perturbation, is maximized when the planet is 
in this region and thus this region is often called as `lensing zone' 
\citep{gould92}.  As the separation departs from the Einstein radius, 
the caustic becomes smaller as $\propto s_{\rm p}^{-2}$ for $s_{\rm p}
\gg 1.0$ and $\propto s_{\rm p}^{2}$ for $s_{\rm p}\ll 1.0$.  In 
addition, the caustic size becomes smaller with the decrease of the 
planet/star mass ratio as $\propto q_{\rm p}^{1/2}$ \citep{han06}.  
When the perturbation is produced by a planet located outside of 
the Einstein ring, the sign of the resulting anomaly in the lensing 
light curve is positive, implying that the magnification during the 
perturbation is higher than the corresponding magnification of the 
single lens event.  On the other hand, if the perturbation is 
produced by a planet located outside of the ring, the sign of the 
anomaly is negative \citep{han03}.

\begin{figure}[t]
\epsscale{1.18}
\plotone{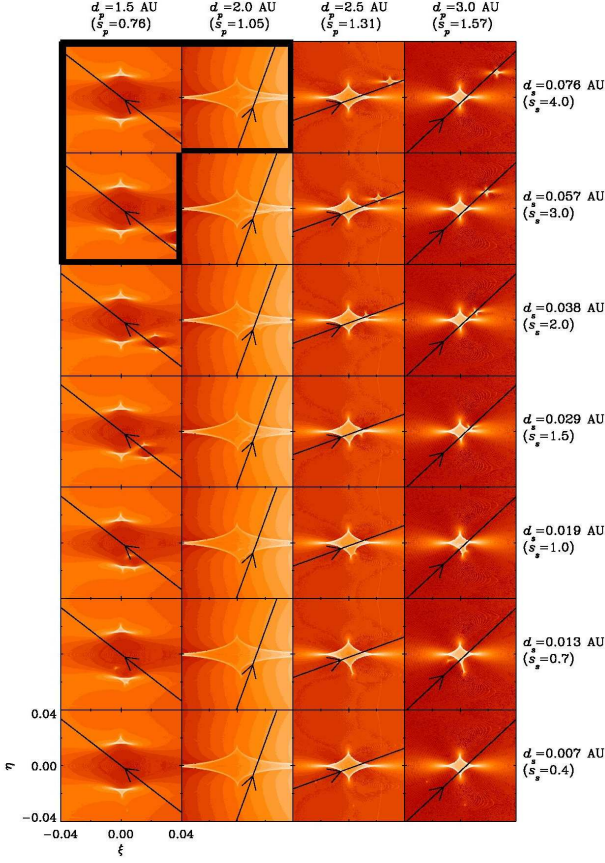}
\caption{\label{fig:one}
Magnification patterns of lens systems composed of a star, a 
planet, and a moon.  The masses of the individual lens components 
are $0.3\ M_\odot$ for the star, $10\ M_{\rm E}$ for the planet, 
and $1.0\ M_{\rm E}$ for the moon. Each map is centered at the 
center of the planetary caustic produced by the planet. The moons 
have a common position angle of $\phi=60^\circ$ with respect to 
the star-planet axis, where the planet is located on the left. 
See Figure~\ref{fig:two} for the geometry of the lens system.  
The labels above and on the right side represent the projected 
star-planet and planet-moon separations, respectively.  The value 
in the parenthesis $s_{\rm p}$ represents the star-planet separation 
in units of the Einstein radius corresponding to the mass of the 
host star, while $s_{\rm s}$ represents the planet-moon separation 
normalized by the Einstein radius corresponding to the mass of the 
planet.  Grey-scale is drawn such that brighter tone represents 
higher magnification.  The panels blocked by thick solid lines 
represent the cases where the planet-moon separation is greater 
than the angle-average of the plant's Hill radius.  The light 
curves resulting from the source trajectories marked by straight 
lines with arrows in the individual panels are presented in the 
corresponding panels of Fig.~\ref{fig:three}.
}\end{figure}

\section{Magnification pattern}

We investigate the microlensing signals of an Earth-like moon 
around ice-giant planets.  For this investigation, we construct 
magnification patterns of lens systems with physical parameters 
adopted from those of typical galactic microlensing events 
currently being detected toward the galactic bulge direction 
\citep{sumi03, udalski03}.  We assume that the planet-hosting 
star is located at a distance of $D_{\rm L}=6\ {\rm kpc}$ from 
the observer and has a mass of $M_\star =0.3\ M_\odot$. We also 
assume that the source star is located at $D_{\rm S} =8\ {\rm kpc}$, 
that corresponds to the distance to the Galactic center.  Then 
the physical Einstein radius corresponding to the lens mass and 
distance is $r_{\rm E}= D_{\rm L}\theta_{\rm E}=1.9$ AU. For the 
star-planet and planet-moon separations, we test various combinations 
keeping in mind that the planet-moon separation should have an upper 
limit. This upper limit is usually set by the Hill radius which 
approximates the gravitational sphere of influence of the planet in 
the face of the perturbation from the host star. The Hill radius is 
related to the semi-major axis, $a$, of the planet and the masses 
of the star, $M_\star$, and planet, $m_{\rm p}$, by
\begin{equation}
r_{\rm H}=a\left( {m_{\rm p}\over 3 M_\star} \right)^{1/3}.
\label{eq5}
\end{equation}
Microlensing is only sensitive to the projected separation, while 
the 3-dimensional separation is important for the orbital stability.
We, therefore, set the angle-average of the projected Hill radius, 
i.e.\ $\sqrt{2/3} r_{\rm H}$, as the upper limit of the planet-moon 
separation.  Since the satellite signal is an additional perturbation 
to the planet-induced perturbation, moons would be detected for 
events where planets are detected.  We, therefore, test planets 
located within the lensing zone of the host star.  In physical 
units, this corresponds to $1.2\ {\rm AU} \lesssim d_{\rm p}\lesssim 
3.0\ {\rm AU}$, where $d_{\rm p}$ is the projected star-planet 
separation.

\begin{figure}[t]
\epsscale{1.18}
\plotone{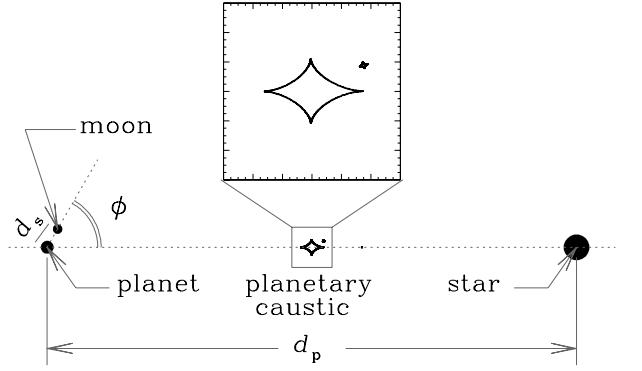}
\caption{\label{fig:two}
Geometry of the lens system composed of a star, a planet, and 
a moon.  The area in the box represents the region where the 
magnification pattern is presented in Figure~\ref{fig:one}.
}\end{figure}

\begin{figure}[t]
\epsscale{1.18}
\plotone{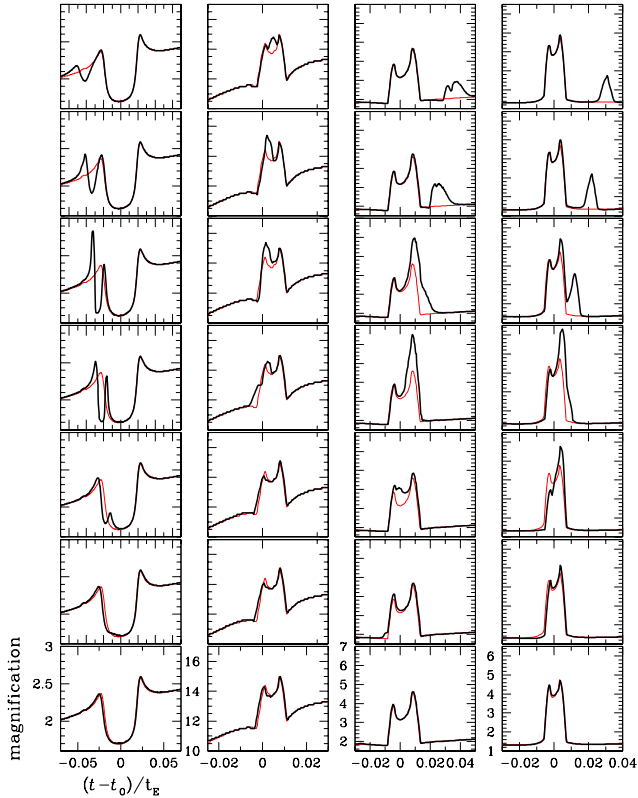}
\caption{\label{fig:three}
Light curves of lensing events produced by lens systems composed 
of a star, a planet, and a moon.  The lens system geometry and 
source trajectories responsible for the individual events are 
presented in the corresponding panels of Fig.~\ref{fig:one}.
In each panel, the thick and thin curves represent the light
curves resulting from lens systems with and without the moon, 
respectively.
}\end{figure}

In Figure~\ref{fig:one}, we present the magnification patterns 
induced by a planet with moons of various projected separations
from the planet.  The planet has a mass of $10\ M_{\rm E}$, 
where $M_{\rm E}$ is the mass of the Earth.  Each map is centered 
at the center of the planetary caustic produced by the planet. 
The moons have a common position angle of $\phi=60^\circ$ with 
respect to the star-planet axis, where the planet is located on 
the left. See Figure~\ref{fig:two} for the geometry of the lens 
system.  The labels above and on the right side of the maps 
represent the projected star-planet and planet-moon separations, 
respectively.  The notations $d_{\rm s}$ and $s_{\rm s}$ represent 
the projected planet-moon separations expressed in physical units 
and in units of the Einstein radius corresponding to the mass of 
the planet, $r_{\rm E,p}$, respectively.  Grey-scale is drawn 
such that brighter tone represents higher magnification.  The 
panels blocked by thick solid lines represent the cases where the 
planet-moon separation is greater than the angle-average of the 
projected Hill radius of the planet and thus moons are prohibited 
to reside.  The light curves resulting from the source trajectories 
marked by straight lines with arrows in the individual panels are 
presented in the corresponding panels of Figure~\ref{fig:three}.  
For the construction of light curves, we take the finite-source 
effect into consideration by assuming that the source star has 
a radius of $1.0\ R_\odot$.

From the magnification patterns and light curves, we find that 
non-negligible satellite signals occur when the planet-moon 
separation is similar to or greater than the Einstein radius of 
the planet, i.e.\ $s_{\rm s}\gtrsim 1.0$.  One thing to be noted 
is that the satellite signal does not diminish with the increase 
of the planet-moon separation beyond the Einstein radius of the 
planet.  This contrasts to the planetary signal that vanishes 
when the planet is located well beyond the Einstein radius of the 
star.  This is because although the lensing effect of the planet 
on the moon rapidly decreases with the increase of the planet-moon 
separation beyond $r_{\rm E,p}$, the effect of the star on the 
moon remains.  When, the planet-moon separation is substantially 
larger than the Einstein radius of the planet, the moon-induced 
perturbation forms at a separate region from the planet-induced 
perturbation region.  In this case, the satellite signal on the 
light curve appears as a separate anomaly and thus it would be 
easily noticed.  When $s_{\rm s}\sim 1.0$, the perturbations 
induced by the planet and satellite interfere each other, 
resulting in complex magnification patterns.  Then, although it 
would be still possible to notice the satellite signal, it would 
be sometimes difficult to unambiguously identify the satellite 
signal.  When the separation is substantially smaller than the 
planetary Einstein radius, the planet and moon behave as if they 
are a single component.  In this case, it would be difficult to 
notice the satellite signal.

Another interesting trend of the satellite signal is that it tends 
to have the same sign as that of the planetary signal.  This trend 
occurs because in most cases of identifiable moons with separations 
from the planet of $s_{\rm p}\gtrsim 1.0$, the lensing effect of 
the host star on the moon is bigger than the effect of the planet.  
Then, the sign of the satellite perturbation is mostly determined 
by the star-moon separation.  The star-moon separation is similar 
to the star-planet separation, and thus the sign of the planet-induced 
perturbation is same as that of the planet-induced perturbation.

\begin{figure}[t]
\epsscale{1.18}
\plotone{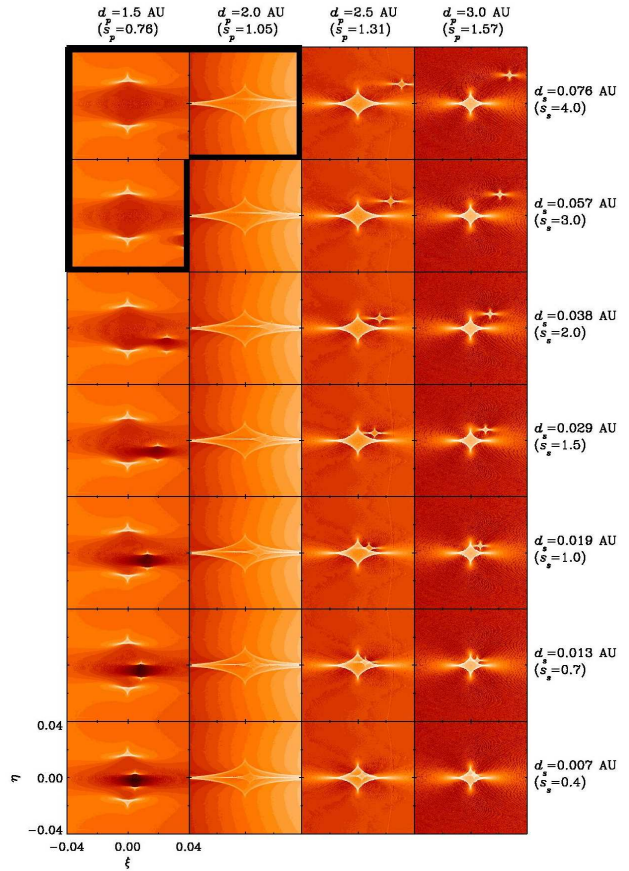}
\caption{\label{fig:four}
Magnification patterns of lens systems obtained by using the binary
superposition approximation. Notations are same as in 
Fig.~\ref{fig:one}.
}\end{figure}

We note that although the planet and moon often reveal themselves 
as separate signals, characterizing them from the independent 
analysis of the individual signals would be difficult. For some cases 
of triple lensing where the effect of the second body on the third 
body is negligible, it is possible to approximate the lensing behavior 
of the triple-lens system as the superposition of the two binary lens 
pairs composed of the first and second bodies and the first and third 
bodies. An example is the multiple-planetary system, where the lensing 
effect of a planet to another planet is negligible \citep{bozza99, 
han01}.  However, for the case of the star-planet-satellite system, 
the effect of the planet on the moon is usually not negligible and 
thus the approximation of binary superposition cannot be used for 
the analysis of the satellite signal.  This can be seen from the 
comparison of the magnification patterns obtained by the exact 
triple-lensing formalism in Figure~\ref{fig:one} and the patterns
obtained by using the binary superposition approximation in 
Figure~\ref{fig:four}.  As expected, it is found that the difference 
in the magnification patterns becomes larger as the planet-moon 
separation decreases.

Then, what will be the range of the satellite separation where the 
microlensing technique is optimized for the detections of moons. 
The lower limit of this range is set by the Einstein radius of the 
planet because moons with separations substantially smaller than 
$r_{\rm E,p}$ are hard to be detected.  Since $r_{\rm E,p}$ depends 
on the planet's mass, planets with different masses have different 
lower limits.  The upper limit is set by the Hill radius because 
moons cannot reside beyond $r_{\rm H}$.  The Hill radius depends 
not only on the planet mass but also on the star-planet separation.  
As a result, even planets with similar masses have different upper 
limits depending on where they are located in the system.

\begin{figure}[t]
\epsscale{1.18}
\plotone{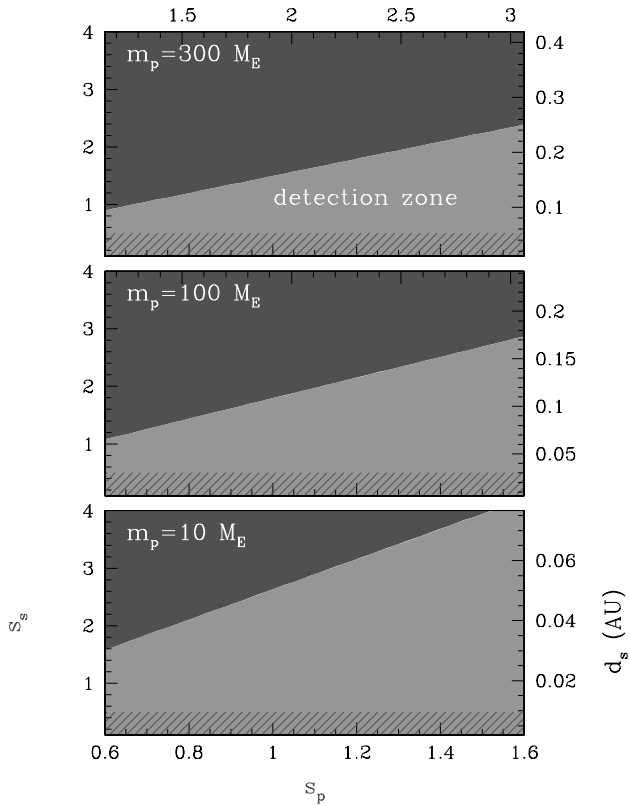}
\caption{\label{fig:five}
The optimal ranges of satellite separations where microlensing
technique is sensitive to the detections of moons.  The labels of 
the star-planet separation on the bottom axis ($s_{\rm p}$) and
the planet-moon separation on the left axis ($s_{\rm s}$) are 
expressed in units of the Einstein radii of the star and planet, 
respectively.  The labels on the top and right axes are expressed 
in physical units.  In each panel, the light-shaded area represents
the region of detectable satellites.  The dark-shade area represents
the region where the planet-moon separation is larger than the 
angle-average of the projected Hill radius of the planet and the 
hatched area represents the region where the separation is smaller 
than half of the planetary Einstein radius.
}\end{figure}

In Figure~\ref{fig:five}, we present the optimal range of satellite 
separations as a function of the star-planet separation for planets 
with different masses.  We note that the labels of the star-planet 
separation on the bottom axis ($s_{\rm p}$) and the planet-moon 
separation on the left axis ($s_{\rm s}$) are expressed in units of 
the Einstein radii of the star and planet, respectively.  The labels 
are expressed also in physical units on the top and right axes, 
respectively.  The individual panels are for planets with masses of 
$m_{\rm p}=300\ M_{\rm E}$, $100\ M_{\rm E}$, and $10\ M_{\rm E}$, 
which roughly corresponds to the masses of Jupiter, Saturn, and 
Uranus, respectively.  In each panel, the light-shaded area represents 
the region of detectable satellites.  The dark-shaded area represents 
the region where the planet-moon separation is larger than the 
angle-average of the projected Hill radius and thus moons are prohibited 
to reside.  The hatched area represents the region where the planet-moon 
separation is smaller than half of the Einstein radius of the planet 
and thus the satellite signal is hard to be detected.  Although the 
range varies depending on the planet's position in the stellar system, 
we find that  the microlensing method would be sensitive to moons 
with separations from the planet of $0.05\ {\rm AU}\lesssim d_{\rm p} 
\lesssim 0.24\ {\rm AU}$ for a Jupiter-mass planet, $0.03\ {\rm AU}
\lesssim d_{\rm p} \lesssim 0.17\ {\rm AU}$ for a Saturn-mass planet, 
and $0.01\ {\rm AU} \lesssim d_{\rm p}\lesssim 0.08\ {\rm AU}$ for 
a Uranus-mass planet.

\begin{figure}[t]
\epsscale{1.18}
\plotone{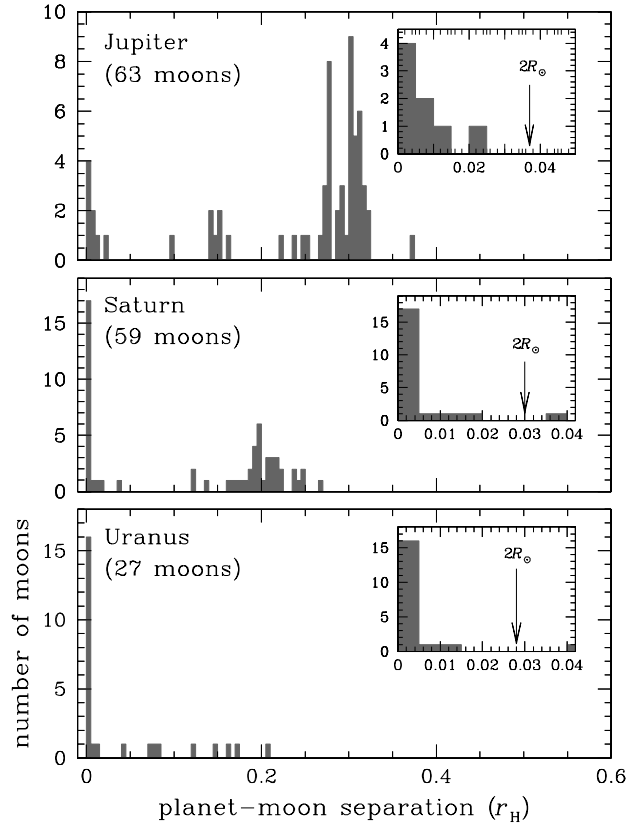}
\caption{\label{fig:six}
The semi-major axis distributions of the moons of Jupiter, Saturn, 
and Uranus.  Note that the planet-moon separations are expressed in 
units of the Hill radii of the individual planets.  The inset in each 
panel shows the distribution for close-in satellites to the planet.  
The arrow represents the upper limit of the planet-satellite separation 
for the detection of moons by using the transit method.
}\end{figure}

\section{Comparison to Transit Method}

Due to the uniqueness of the microlensing method in detecting 
planets and their moons, the characteristics of the moons to be 
detected by the microlensing method will be different from those 
to be discovered by the transit method.  Below, we list some of 
these differences.

First, while the transit method can be used to search for moons of 
nearby stars, the microlensing method are sensitive to moons of 
remote stars.  To meet the precision of photometry that is required 
to detect moons of extrasolar planets, the target stars of transit 
searches should be bright and thus they are confined to the solar 
neighborhood.  By contrast, microlensing searches are sensitive to 
stars anywhere along the line-of-sight toward the Galactic bulge.
Therefore, the microlensing method can provide a sample of extrasolar 
moons distributed throughout the galaxy.

Second, while the transit method is most sensitive to moons of 
close-in planets, the microlensing method is sensitive to moons of
planets in the region beyond the `snow line'.  The snow line is 
the point in the protoplanetary disk beyond which the temperature 
is less than the condensation temperature of water \citep{lecar06}.
Enhanced surface density of solids helps the formation of cores of 
giant planets and thus giant planets are thought to form in the 
region immediately beyond the snow line.  The giant planets in our 
solar system, which are located in this region, have numerous moons; 
63 known moons for Jupiter, 59 for Saturn, and 27 for Uranus.
On the contrary, there might be few moons in close-in planets due 
to the strong tidal effect of the host stars as suggested by the 
two innermost planets of Mercury and Venus in our solar system.

Third, while the transit method is sensitive to moons located close 
to their host planets, moons detectable by the microlensing method 
will have wide separations from the planets.  In order to produce 
additional dips in transit light curves, moons should be located 
very close to their planets with separations equivalent to or less 
than the diameter of the host star.  On the other hand, the sensitivity 
of microlensing method  extends up to the Hill radius.  For the case 
of giant planets in our solar system, the numbers of moons with 
separations larger than $0.1 r_{\rm H}$ are 54 (85.7\% of the 
all known moons), 38 (64.4\%), and 5 (18.5\%) for Jupiter, Saturn, 
and Uranus, respectively.  See the semi-major axis distributions 
of the moons of Jupiter, Saturn, and Uranus in Figure~\ref{fig:six}.  
In these planets, there also exist close-in moons with separations 
from the planets less than the diameter of the sun; 7 (11.1\%), 19 
(32.2\%), and 17 (63.0\%) for Jupiter, Saturn, and Uranus, respectively.  
However, it would be difficult to detect them by using the transit 
method because the star-planet separation is large and thus the 
probability of planet transit is very low.

\section{Conclusion}

We investigated the characteristic of microlensing signals of 
Earth-like moons orbiting ice-giant planets.  For this, we constructed 
magnification patterns of lens systems with various star-planet 
and planet-moon separations.  From this investigation, we found 
that non-negligible satellite signals occur when the planet-moon 
separation is similar to or greater than the Einstein radius of 
the planet.  We found that the satellite signal does not diminish 
with the increase of the planet-moon separation beyond the Einstein 
radius of the planet unlike the planetary signal which vanishes
when the planet is located well beyond the Einstein radius of the 
star.  We also found that the satellite signal tends to have the 
same sign as that of the planetary signal.  These tendencies are 
caused by the lensing effect of the star on the moon in addition 
to the effect of the planet.  We determined the range of satellite 
separations where the microlensing technique is optimized for the 
detections of moons.  By setting an upper limit as the angle-average 
of the projected Hill radius and a lower limit as half of the Einstein 
radius of the planet, we found that the microlensing method would be 
sensitive to moons with projected separations from the planet of 
$0.05\ {\rm AU} \lesssim d_{\rm p} \lesssim 0.24\ {\rm AU}$ for 
a Jupiter-mass planet, $0.03\ {\rm AU}\lesssim d_{\rm p} \lesssim 
0.17\ {\rm AU}$ for a Saturn-mass planet, and $0.01\ {\rm AU} 
\lesssim d_{\rm p} \lesssim 0.08\ {\rm AU}$ for a Uranus-mass 
planet.  We compared the characteristics of the moons to be detected 
by the microlensing and transit techniques.

\acknowledgments 
This work was supported by the Astrophysical Research Center for the
Structure and Evolution of the Cosmos (ARCSEC) of Korea Science and
Engineering Foundation (KOSEF) through Science Research Program (SRC)
program.

\end{document}